\documentclass[final,3p,times]{elsarticle}
\usepackage{amssymb}
\usepackage{amsmath}
\usepackage{amsthm}
\usepackage{lineno}
\usepackage{graphicx}
\usepackage{multirow}
\usepackage{tikz}
\usetikzlibrary{positioning}
\usetikzlibrary{fit,arrows, decorations.markings}
\tikzstyle{vecArrow} = [thick, decoration={markings,mark=at position
	1 with {\arrow[semithick]{open triangle 60}}},
double distance=1.4pt, shorten >= 5.5pt,
preaction = {decorate},
postaction = {draw,line width=1.4pt, white,shorten >= 4.5pt}]
\tikzstyle{sbox} = [draw=black, fill=white!15, very thick,text width=5cm,
rectangle, rounded corners, inner sep=3pt, inner ysep=6pt,align=center]
\usepackage{tabularx}
\biboptions{sort&compress}
\newcounter{bla}

\journal{Computer Physics Communications}
\newcommand{\Longitud}{{\shortparallel}}
\newcommand{\Transvers}{{\perp}}

\begin{document}
\begin{frontmatter}
\title{Polarized NLO EW $e^+e^-$ cross section calculations with {\tt ReneSANCe-v1.0.0}}

\author[a]{Renat~Sadykov\corref{author}}
\ead{sadykov@cern.ch}
\author[a,b]{Vitaly~Yermolchyk\corref{author}}
\ead{Vitaly.Yermolchyk@jinr.ru}

\cortext[author]{Corresponding authors.}
\address[a]{Dzhelepov Laboratory for Nuclear Problems, JINR, Joliot-Curie 6, RU-141980 Dubna, Russia}
\address[b]{Institute for Nuclear Problems, Belarusian State University, Bobruiskaya 11, 220006 Minsk, Belarus}

\begin{abstract}

In this paper we present a new Monte Carlo event generator {\tt ReneSANCe} for simulation of processes at electron-positron colliders. In the current release of the generator the Bhabha scattering
($e^+e^- \to e^-e^+$) and Higgs-strahlung ($e^+e^- \to ZH$) process are implemented. Based on the SANC (Support for Analytic and Numeric Calculations for experiments at colliders) modules, the new
generator takes into account complete one-loop and some higher-order electroweak radiative corrections with finite particle masses and polarizations. The new generator effectively operates in the
collinear region and at the $ZH$ production threshold. It is constructed in such a way that new processes can be easily added. The paper contains a theoretical description of the SANC approach, numerical
validations and manual.

\end{abstract}

\begin{keyword}

Perturbation theory; NLO calculations; Standard Model; Electroweak interaction; QED; Monte Carlo simulation

\end{keyword}

\end{frontmatter}

{\bf PROGRAM SUMMARY}

\begin{small}
\noindent
{\em Program Title: ReneSANCe-v1.00}\\
{\em Licensing provisions: GPLv3}\\
{\em Programming language: Fortran, C, C++}\\
{\em Supplementary material: Looptools~[1], FOAM~[2]}\\
{\em Nature of problem: Theoretical calculations at next-to-leading order in
perturbation theory allow to compute higher precision amplitudes for Standard
Model processes and decays, provided proper treatments of UV divergences and IR
singularities are performed}\\
{\em Solution method: Numerical integration of the precomputed differential expressions for
cross sections of certain processes implemented as SANC modules~[3,4]}\\
{\em Restrictions: the list of processes is limited to $e^+e^- \to e^-e^+$ and $e^+e^- \to ZH$}\\

\end{small}

\section{Introduction}
\label{sec:introduction}
The construction of high-precision theoretical predictions and their comparison with experimental data play a crucial role in solving the problem of applicability of the Standard Model
and justifying its structure on the basis of fundamental principles. Implementation of the results of theoretical calculations in Monte Carlo event generators is an important
step of theoretical support for high-precision experimental verification of the Standard Model carried out in modern and future experiments. 

Monte Carlo event generators are used to account for detector effects in experimental data and obtain predictions with which these experimental data will be compared. In addition, they can be used to
obtain histograms of complex observables and pseudo-observables without rerunning and rewriting the code. One can simply take the generated events and analyze them using programs
such  as {\tt RIVET} \cite{Buckley:2010ar} or {\tt ROOT} \cite{Brun:1997pa}, in contrast to integrators where histograms are often hardcoded.

It is very important to comprehensively take into account the effects of higher orders
corrections due to strong, electromagnetic and weak interactions. An additional option of existing
theoretical prediction tools should be the consideration of the beam polarization of future electron-positron accelerators. Provided that both beams are polarized, experimental tests can be performed with unprecedented
accuracy, either at the $Z$ pole, or at the $WW$ threshold, or at the peak of the $ZH$ process, as well as at the $t\bar{t}$ threshold. Accounting for polarization will have far-reaching consequences for studies
on the consistency of the electroweak theory, in particular in the Higgs sector.

This paper describes the Monte Carlo event generator {\tt ReneSANCe} (\textbf{Rene}wed \textbf{SANC}
Monte Carlo \textbf{e}vent generator), which provides a next-to-leading order (NLO) accurate electroweak (EW)
description of some important processes at electron-positron colliders with respect
to polarization effects. In the current release of the generator the Bhabha scattering ($e^+e^- \to e^-e^+$) \cite{Bardin:2017mdd} and Higgs-strahlung ($e^+e^- \to ZH$) process \cite{Bondarenko:2018sgg} are implemented.

Combination of NLO EW corrections with the effect of polarization of initial particles allows extraction of the Standard Model parameters with greatly improved precision.

At the tree level, the processes of $e^+e^-$ collisions with polarization of initial particles are realized in the Monte Carlo programs {\tt CalcHEP} \cite{Belyaev:2012qa} and {\tt WHIZARD} \cite{Kilian:2007gr}.

Summing up, there is a necessity for a program code implemented as a generator and
that takes into account electroweak corrections and corrections of higher orders, with the possibility of expanding the list of processes available for calculation.

The paper is organized as follows. Section~\ref{sec:physics} contains a brief description
of one-loop electroweak calculations in the {\tt SANC} system taking into account the polarization
of initial particles. The structure of the code is described in section~\ref{sec:code}.
The benchmarks against existing tools and calculations are presented in
section~\ref{sec:validation}. Summary is given in section~\ref{sec:summary}.

\section{NLO EW corrections with polarization in {\tt SANC} framework}
\label{sec:physics}
The calculations are organized in a way that allows one to control the consistency of the result. All analytical calculations at the one-loop precision level are realized in the $R_\xi$ gauge with three gauge
parameters: $\xi_{\mathsf{A}}$, $\xi_{\mathsf{Z}}$, and $\xi\equiv\xi_{\mathsf{W}}$. To parameterize ultraviolet divergences, dimensional regularization is used. Loop integrals are expressed in
terms of the standard scalar Passarino-Veltman functions: $ A_0, \, B_0, \, C_0, \, D_0$ \cite{PASSARINO1979151}. These features make it possible to carry out several important checks at the level of analytical expressions,
e.g., checking the gauge invariance by eliminating the dependence on the gauge parameter, checking cancellation of ultraviolet poles, as well as checking various symmetry properties and
the Ward identities.

\begin{figure}[!ht]
\centering
\includegraphics[]{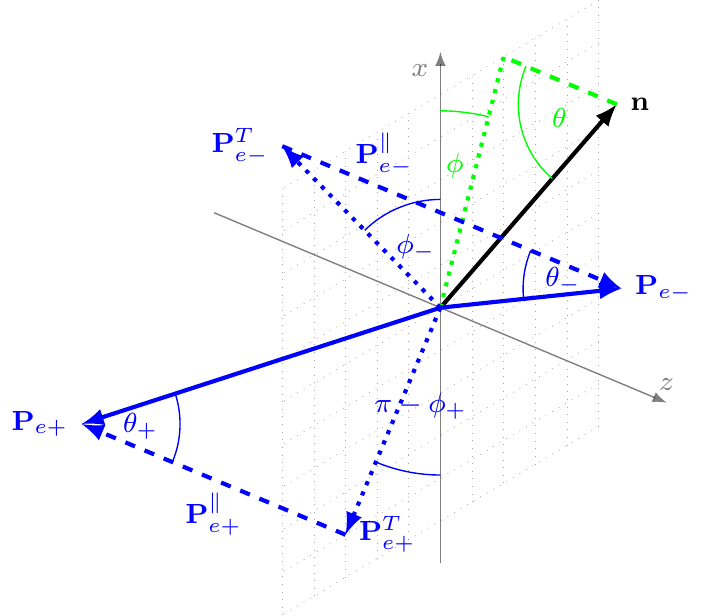}
\caption{
Decomposition of the $e^{\pm}$ polarization vectors.}
\label{fig:polar}
\end{figure}

Polarization states of initial positron and electron beams can be described by the polarization vectors $\vec{P}_{e^+}$ and $\vec{P}_{e^-}$, respectively. The magnitude of these vectors represents the
polarization degree and varies from 0 to 1. We can decompose these vectors into longitudinal $\vec{P}_{e^\pm}^{\Longitud}$ and transverse $\vec{P}_{e^\pm}^{\Transvers}$ parts, as illustrated in
Fig.~\ref{fig:polar}.

Using helicity amplitudes, we can calculate the matrix element squared of the generic process $e^+e^- \to X$ for an arbitrary polarization state of the positron and electron
beams \cite{MoortgatPick:2005cw}:
\begin{eqnarray}
|\mathcal{M}|^2 &= &L_{e^+}^{\Longitud}R_{e^-}^{\Longitud}|\mathcal{H}_{-+}|^2
+R_{e^+}^{\Longitud} L_{e^-}^{\Longitud}|\mathcal{H}_{+-}|^2
+L_{e^+}^{\Longitud} L_{e^-}^{\Longitud}|\mathcal{H}_{--}|^2
+R_{e^+}^{\Longitud} R_{e^-}^{\Longitud}|\mathcal{H}_{++}|^2
\nonumber\\
&&-\frac{1}{2}P_{e^+}^{\Transvers} P_{e^-}^{\Transvers} \Re\Bigl[e^{i(\Phi_+-\Phi_-)}
\mathcal{H}_{++}\mathcal{H}_{--}^{*}
+e^{i(\Phi_++\Phi_-)}\mathcal{H}_{-+}\mathcal{H}_{+-}^{*}\Bigr]
\nonumber\\
&&+P_{e^-}^{\Transvers} \Re\Bigl[e^{i\Phi_-}\Bigl(L_{e^+}^{\Longitud}
\mathcal{H}_{-+}\mathcal{H}_{--}^{*}
+R_{e^+}^{\Longitud}\mathcal{H}_{++}\mathcal{H}_{+-}^{*}\Bigr)\Bigr]
\nonumber\\
&&-P_{e^+}^{\Transvers} \Re\Bigl[e^{i\Phi_+}\Bigl(L_{e^-}^{\Longitud}
\mathcal{H}_{+-}\mathcal{H}_{--}^{*}
+R_{e^-}^{\Longitud}\mathcal{H}_{++}\mathcal{H}_{-+}^{*}\Bigr)\Bigr],
\label{polxsec}
\end{eqnarray}
where
\begin{equation}
L_{e^\pm}^{\Longitud} = \frac{1}{2}(1-P_{e^\pm}^{\Longitud}), \quad R_{e^\pm}^{\Longitud} = \frac{1}{2}(1+P_{e^\pm}^{\Longitud}),
\quad \Phi_\pm = \phi_\pm - \phi,
\nonumber
\end{equation}
$\mathcal{H}_{--}$, $\mathcal{H}_{++}$, $\mathcal{H}_{-+}$, $\mathcal{H}_{+-}$
are helicity amplitudes with +(-) subscript corresponding to the case when spin of the particle is parallel (antiparallel) to momentum of that particle. Here the first subscript refers to the positron,
and the second one to the electron.

In the current version of {\tt ReneSANCe} we implemented only the longitudinal polarization of initial particles.

The cross section of the process $e^+e^- \to X$ at the one-loop EW level can be divided into four parts:
\begin{eqnarray}
\sigma^{\text{1-loop}} = \sigma^{\mathrm{Born}} + \sigma^{\mathrm{virt}}(\lambda) + \sigma^{\mathrm{soft}}(\lambda, \omega)
+ \sigma^{\mathrm{hard}}(\omega),
\label{loopxsec}
\end{eqnarray}
where $\sigma^{\mathrm{Born}}$ is the Born level cross section, $\sigma^{\mathrm{virt}}$ is a contribution of virtual (loop) corrections, $\sigma^{\mathrm{soft}}$ corresponds to a soft photon
emission, and $\sigma^{\mathrm{hard}}$ is a hard photon emission part (with energy $E_{\gamma} > \omega$). Auxiliary parameters $\lambda$ (fictitious "photon mass" which regularizes infrared divergences) and
$\omega$ (photon energy which separates the regions of the phase space associated with the soft and hard emissions) cancel out after summation.

We use the helicity approach for all contributions.
It provides us the possibility to describe in the future any initial (not only longitudinal) polarization, polarization of final states, spin correlations, polarization transfer from initial to final states. In {\tt ReneSANCe-v1.0.0} the helicity approach is fully
implemented for the Higgs-strahlung  process, but for the Bhabha scattering $\sigma^{\mathrm{hard}}$ is calculated using the standard covariant amplitude approach.

Calculations were carried out in the {\tt SANC} framework \cite{ANDONOV2006481} using {\tt FORM} \cite{Vermaseren:2000nd}. For optimization we intensively used factorization \cite{Kuipers:2012rf} introduced in {\tt FORM} 4  and elimination
of common subexpressions with temporary variables \cite{Kuipers:2013pba}. The results of calculations were automatically transformed into software modules in the {\tt FORTRAN} 77 language with the standard {\tt SANC} interface.
Evaluation of loop integrals in the modules is performed by {\tt Looptools} \cite{Hahn:1998yk} and {\tt SANClib} \cite{Bardin:2009zz,Bardin:2009ix} packages.
These modules incorporating physical calculations can be used in any code that understands the {\tt SANC} interface \cite{Andonov:2008ga}.

\begin{figure}[!h]
	\begin{center}
		\includegraphics[width = 0.45\textwidth]{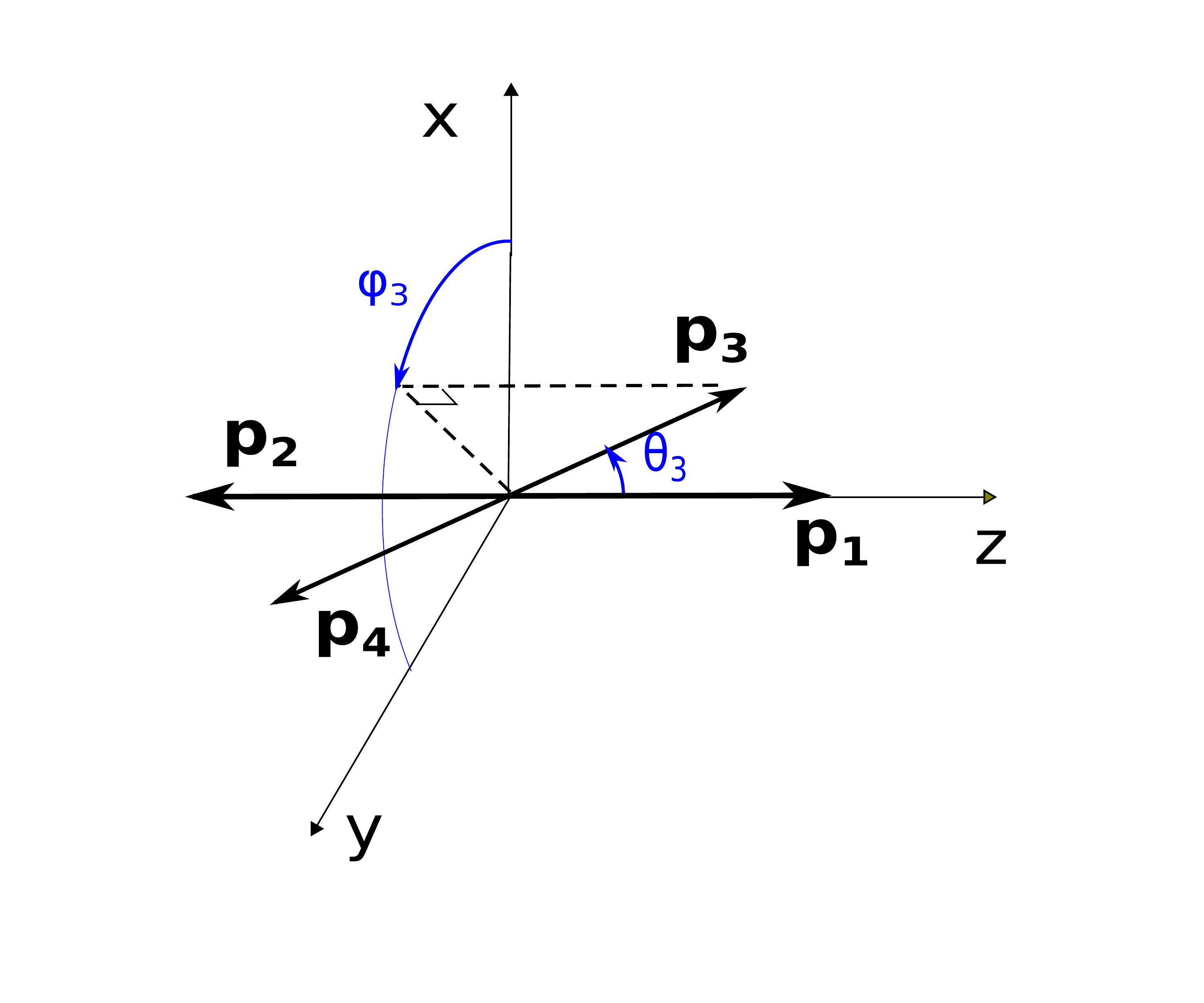}
		\includegraphics[width = 0.45\textwidth]{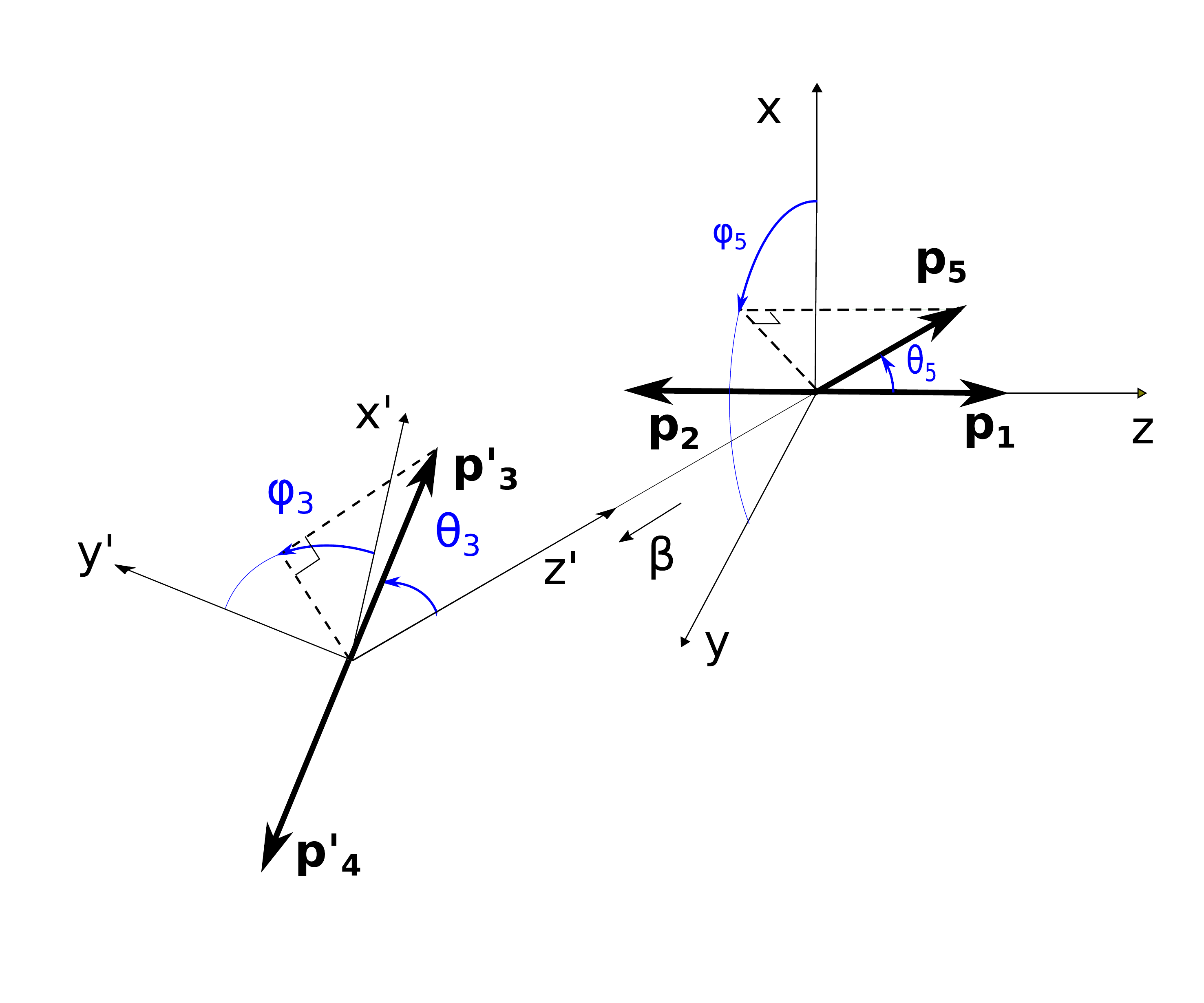}
	\end{center}
	\caption{Kinematics of the $2 \to 2$ (left) and $2 \to 3$ (right) processes.
		3-momenta $\vec{p}_1$, $\vec{p}_2$ and $\vec{p}_5$ lie in the $y'z'$-plane,
		$\beta = \dfrac{|\vec{p}_3+\vec{p}_4|}{E_3+E_4} = \dfrac{s-s'}{s+s'}$.}
	\label{fig:kinematics}
\end{figure}

We used the following parametrization of the phase space (Fig.~\ref{fig:kinematics}).
\begin{itemize}
	\item The phase space of the $2 \to 2$ process is parameterized via
	\begin{itemize}
		\item $\theta_3$ --- the angle
		between $\vec{p}_3$ and $\vec{p}_1$ in the center-of-momentum frame of particles 1 and 2;
		\item $\phi_3$ --- the azimuthal angle
		of particle 3 with respect to $\vec{p}_1$.
	\end{itemize}
	\item The phase space of the $2 \to 3$ process is parameterized via
	\begin{itemize}
		\item $\theta_5$ --- the angle
		between $\vec{p}_5$ and $\vec{p}_1$ in the center-of-momentum frame of particles 1 and 2;
		\item $\phi_5$ --- the azimuthal angle of particle 5 with respect to $\vec{p}_1$;
		\item $\theta_3$ --- the angle
		between $\vec{p}_3$ and $\vec{p}_5$ in the center-of-momentum frame of particles 3 and 4;
		\item $\phi_3$ --- the azimuthal angle of particle 3 with respect to $\vec{p}_5$;
		\item $s'$ --- the square of the invariant mass of particle 3 and 4: $s' = (p_3+p_4)^2$.
	\end{itemize}
\end{itemize}
This parametrization is compatible with the one used in {\tt SANC} modules.
In these variables the cross sections are given by the following phase-space integrals:
\begin{equation*}
\sigma(2\to2) = \int\limits_{-1}^{1}d\cos{\theta_3}\int\limits_{0}^{2\pi}d\phi_3
\dfrac{1}{64\pi^2s}\dfrac{\sqrt{\lambda(s',m_3^2,m_4^2)}}{\sqrt{\lambda(s,m_1^2,m_2^2)}}|\mathcal{M}_{2}|^2 F_2.
\end{equation*}
for the 2$\to$2 processes and
\begin{eqnarray*}
\sigma(2\to3)
= \int\limits_{(m_3+m_4)^2}^{s(1-\omega)}ds'\int\limits_{-1}^{1}d\cos{\theta_3}\int\limits_{0}^{2\pi}d\phi_3\int\limits_{-1}^{1}d\cos{\theta_5}\int\limits_{0}^{2\pi}d\phi_5
\dfrac{s-s'}{4096\pi^5ss'}\dfrac{\sqrt{\lambda(s',m_3^2,m_4^2)}}{\sqrt{\lambda(s,m_1^2,m_2^2)}}|\mathcal{M}_{3}|^2 F_3.
\end{eqnarray*}
for  the 2$\to$3 processes. Here $F_2 = F_3 = 0$ for the region of phase-space excluded by kinematic cuts, and $F_2 = F_3 = 1$ otherwise.

According to the {\tt SANC} agreement, trivial integration over $\phi_3$ for the $2 \to 2$ processes and over $\phi_5$ for the $2 \to 3$ processes is performed in the corresponding module for the differential cross section,
so we need to take it into account to prevent double counting.

\section{{\tt ReneSANCe} structure}
\label{sec:code}

\begin{figure}[!ht]
	\centering
\begin{tikzpicture}
\node[sbox,text width=4cm](feyngen){cl Settings};
\node[sbox,right= of feyngen](feyngenComm){read/validate preferences};
\node[sbox](Graphine)[below = of feyngen]{cl Info};
\node[sbox,right= of Graphine](GraphineComm){calculate derived parameters, store results};
\node[sbox][below = of Graphine](PyGraphState1){\texttt{libSANC}, \texttt{SANC} modules};
\node[sbox,right= of PyGraphState1](PyGraphState1Comm){calculate matrix element};
\node[sbox][below = of PyGraphState1](PyGraphState){\texttt{ROOT::mFOAM}};
\node[sbox,right= of PyGraphState](PyGraphStateComm){grid construction \& phase-space sampling};
\node[sbox][below = of PyGraphState](Mako){cl DataSaver, utilities};
\node[sbox][right= of Mako](MakoComm){write events: root, LHEF, ...\\ \& analyse};
\draw [-] (feyngen) to (feyngenComm);
\draw [vecArrow] (feyngen) to (Graphine);
\draw [-] (GraphineComm) to (Graphine);
\draw [-] (PyGraphState1Comm) to (PyGraphState1);
\draw [-] (PyGraphStateComm) to (PyGraphState);
\draw [vecArrow] (Graphine) to (PyGraphState1);
\draw [vecArrow] (PyGraphState1) to (PyGraphState);
\draw [vecArrow] (PyGraphState) to (Mako);
\draw [-] (MakoComm) to (Mako);
\end{tikzpicture}
\caption{
	Simplified scheme of {\tt ReneSANCe} event generator structure.}
\label{fig:struct}
\end{figure}
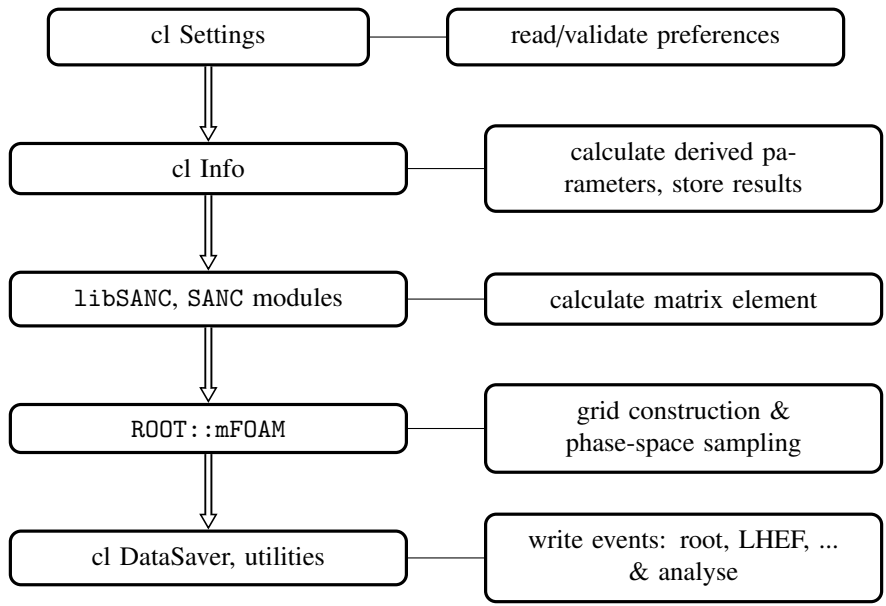

\begin{figure}[!ht]
\centering\fbox{\includegraphics[width=12cm,trim={0 3.2cm 0 0},clip]{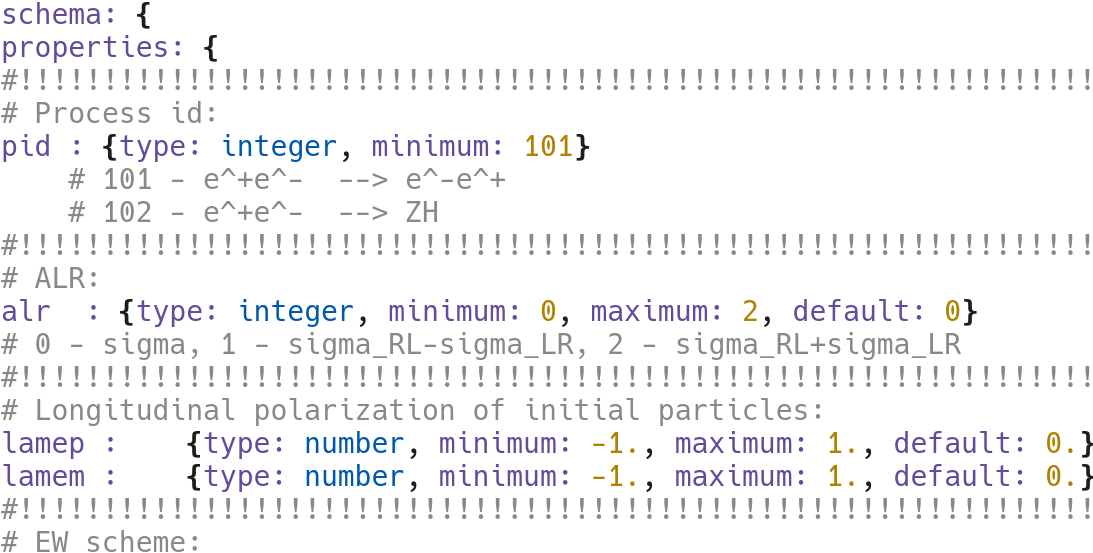}}
\caption{
	{\tt ReneSANCe} schema file used for validation of input parameters.}
\label{fig:schema}
\end{figure}

\begin{figure}[!ht]
\centering\fbox{\includegraphics[width=12cm,trim={0 3.2cm 0 0cm},clip]{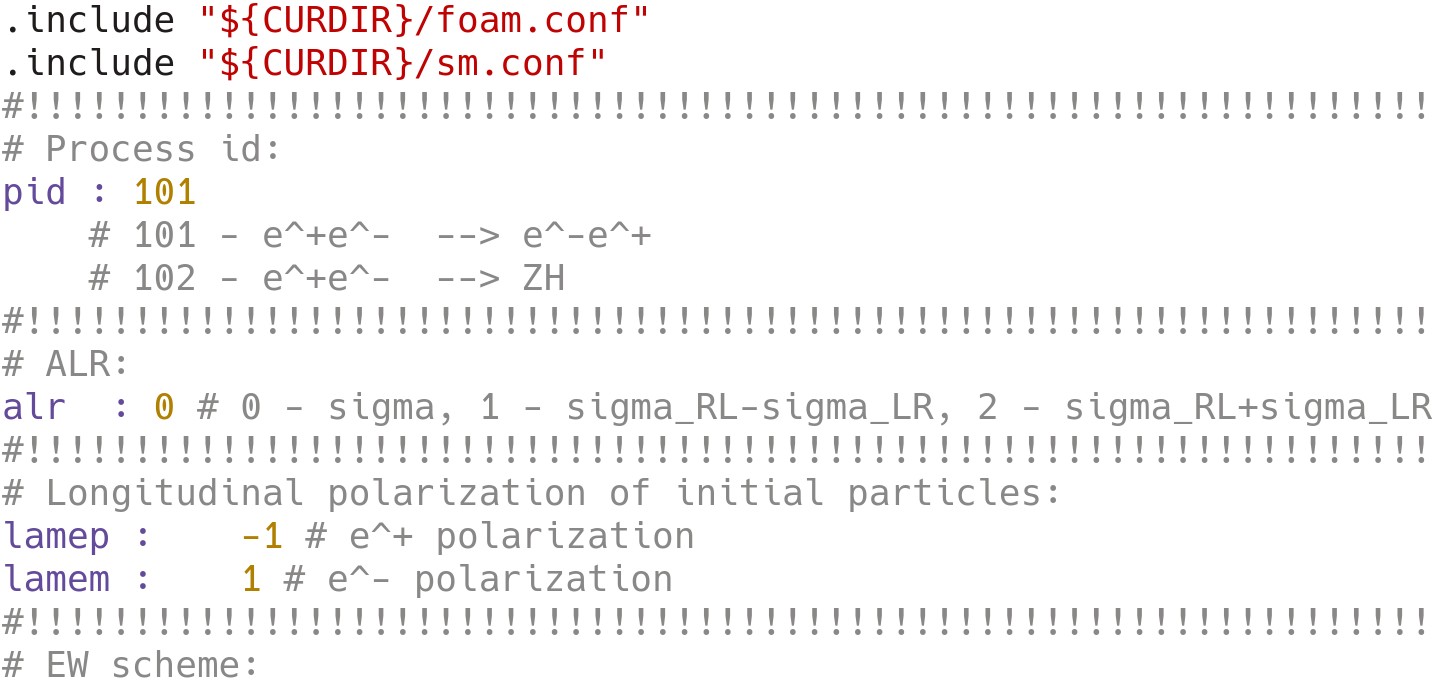}}
\caption{
	Example of {\tt ReneSANCe} settings file.}
\label{fig:settings}
\end{figure}

\begin{figure}[!ht]
	\centering\includegraphics[width=8cm]{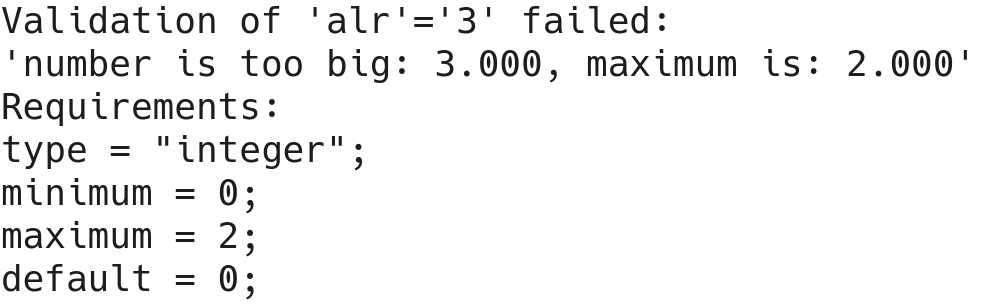}	
\caption{
	Example of the validation subsystem response.}
\label{fig:validation}
\end{figure}

The {\tt ReneSANCe} is written as a set of modules (Fig.~\ref{fig:struct}).
In the current realization we use a modified version of 
{\tt libucl} \cite{libucl} for work with settings. All actual settings are stored internally as a 'key':'value'
map. Default settings from the schema files, settings from the user input file 
(Fig.~\ref{fig:settings}) and the received through command line interface are merged together according to priorities (the lowest for default one and the highest for command line). 
The work with command line is done using header only library {\tt CLI11} \cite{cli11}.
After all the settings are merged together, they are checked to be valid according
to the schema files (Fig.~\ref{fig:schema}). If any error is found, the program is terminated with the corresponding error
message (Fig.~\ref{fig:validation}).
As the next step, the Info class calculates all the derived parameters and initializes {\tt SANC} modules.

For generating unweighted events, according to the differential cross section, and calculating the total integral we use an adaptive algorithm {\tt mFOAM} \cite{Jadach:2005ex} that is part of the {\tt ROOT} program.
The event generation procedure can be divided into several stages.
At the first stage, {\tt mFOAM} explores the integrand and produces a grid that divides the whole integration domain into smaller hyperrectangular cells. At the next stage, {\tt mFOAM} uses the grid
to sample over the integration domain, according to the differential cross section, which is provided
by the {\tt SANC} modules written in {\tt FORTRAN} 77.
Then {\tt ReneSANCe} evaluates momenta of all particles and writes 
event in output files. In the first release, we support {\tt LHEF} and {\tt ROOT} output
formats.

For sampling we use a multibranching strategy with variable transformation. We implemented two approaches.
The first (``manual'' approach) is based on manual sampling over branches associated with different parts of one-loop cross section given in Eq. \ref{loopxsec}. For each branch we created a separate instance of the {\tt FOAM} class.
In this case, each branch can use both optimal variable transformation and the optimal {\tt FOAM} setup.
As a consequence, an additional stage is needed to calculate branching weights that slow down the initialization stage of the generator.
In the second (``all-in-one'') approach, sampling is made using only one instance of the {\tt FOAM}. Nevertheless, optimal variable transformation for each branch is also available. The {\tt FOAM} is responsible for sampling over branching.
It is performed by creating additional artificial dimension of integral with fixed division points.

\subsection{Installation}
The {\tt ROOT} framework must be available at the stage of compilation.
There are several opportunities for its installation, e.g., from distributive repository, from the official site or activate it from CernVM-FS repositories located under /cvmfs.
For example, in the case of CERN Scientific Linux 6 (SLC6) with gcc 4.9.* compiler in 64-bit mode 
\begin{verbatim}
source /cvmfs/sft.cern.ch/lcg/views/ROOT-latest/x86\_64-slc6-gcc49-opt/setup.sh 
\end{verbatim}

Other dependencies are bundled into archive with {\tt ReneSANCe}.

The {\tt ReneSANCe} event generator uses the {\tt CMake} build system.
To install {\tt ReneSANCe} run
\begin{verbatim}
	cmake <path_to_source> -DCMAKE_INSTALL_PREFIX=<inst_prefix> <other options>
	cmake --build .
	cmake --build . --target install
\end{verbatim}

To list all available {\tt CMake} options with the corresponding descriptions, one can run
\begin{verbatim}
cmake -LH <path_to_source>
\end{verbatim}

\subsection{Setup}
{\tt ReneSANCe} needs an access to its schema files at run time to work properly.
It searches them relatively to the installation path.
If the generator has been relocated and cannot find schema files, run
\begin{verbatim}
source renesance-init.sh
\end{verbatim}
to make it available in the current console or export environment variable {\tt RENESANCE\_ROOT}.
For bash shell:
\begin{verbatim}
export RENESANCE_ROOT=<path_to_root_directory_of_ReneSANCe_installation>
\end{verbatim}

Program can be run as
\begin{verbatim}
ReneSANCe <option1> <option2> ...
\end{verbatim}

The {\tt ReneSANCe} event generator has a multilayer configuration system: default parameters are defined in the program, parameters are provided by a user through configuration files (higher priority)
and command line parameters (the highest priority).

Many parameters would be initialized to a default value, which the user can find in the schema files installed in system.

However, some parameters like `pid', `ecm' do not have default value, so the user must provide it manually using a configuration file or command line interface.

The list of command line options:
\begin{verbatim}
-h,--help                   Print this help message and exit
-f,--file FILE              Set path to index.conf file
-s,--seed INT               Set seed
-p,--pid INT in [101 - 102] Set process
-e,--ecm FLOAT              Set energy of collider
--pol1 FLOAT in [-1 - 1]    Set first beam polarization
--pol2 FLOAT in [-1 - 1]    Set second beam polarization
-D,--define TEXT ...        Set other settings as list of key:value
\end{verbatim}

By default the program searches for a configuration file that has the name \textit{index.conf} in the current directory. However, it can be provided by the {\tt ReneSANCe} using the command line parameter `-f'.

The input file language was built on JSON syntax, so configuration in JSON format is valid.
Our input format is less restrictive. The line comments are allowed with \# as a separator.
Also, opening and closing brackets \{ \} can be omitted.
The parameters are initialized using the key:value or key=value syntax.
The type of each key is defined in the schema files.

\subsection{Configuration}

Configuration files support JSON-like syntax and contain general steering parameters for a run, {\tt FOAM}  parameters, kinematic cuts and parameters of the Standard Model.
The configuration can be split into several files and included to index.conf file by \textit{.include $<$relative or absolute path to file$>$} macro.

\subsubsection{Process parameters}
\begin{description}
	\item [pid] [{\tt integer}] = defines a process to calculate (\texttt{integer})\\
		101: $e^+e^- \to e^-e^+$\\
		102: $e^+e^- \to ZH$

	\item [ecm] [{\tt double}] sets collider energy in the center-of-momentum frame,
	\item [alr] [{\tt integer}] = switch generation mode\\
		0: $\sigma$\\
		1: $\sigma_{RL}-\sigma_{LR}$\\
		2: $\sigma_{RL}+\sigma_{LR}$
	\item [lamep] [{\tt double}] polarization degree of the positron,
	\item [lamem] [{\tt double}] polarization degree of the electron,
	\item [costhcut] [{\tt double}] cut on $|\cos\theta|$ for both final state particles,
	\item [ome] [{\tt double}] is the parameter separating contributions from soft and hard photon Bremsstrahlung ($\omega=ome \sqrt{s}/2$).
\end{description}
	Flags controlling components of the NLO EW computations:
	\begin{description}
		\item [iqed] [{\tt integer}] = \hfill \\
		0: disables QED corrections\\
		1: with full QED corrections\\
		2: only initial state QED radiation (ISR)\\
		3: initial-final QED radiation interference term (IFI) \\
		4: only final state QED radiation (FSR)\\
		5: sum of initial and final state radiation contributions [IFI+FSR]\\
		6: sum of initial state and initial-final QED interference terms [ISR+IFI]
		\item [iew] [{\tt integer}] = 0/1 corresponds to disabled/enabled weak corrections,
		\item [iborn] [{\tt integer}] = 0 or 1 selects respectively LO or NLO level of calculations,
		\item [ifgg] [{\tt integer}] = choice of calculations for photonic vacuum polarization \(\mathcal{F}_{\gamma\gamma}\)\\
		-1: 0 \\
		0: 1 \\
		1: \(1+\mathcal{F}_{\gamma\gamma}(\mathrm{NLO})\) \\
		2: \(1/[1 - \mathcal{F}_{\gamma\gamma}(\mathrm{NLO})]\)
		\item [irun] [{\tt integer}] = 0/1 corresponds to fixed/running gauge boson width,
		\item [gfscheme] [{\tt integer}] = flag selects electroweak scheme in which the calculation is performed\\
		0: $\alpha(0)$-scheme\\
		1: $G_{\mu}$-scheme
	\end{description}

\subsubsection{FOAM parameters}
By default we use optimal {\tt FOAM} parameters, but a user can tune some parameters manually:
\begin{description}
	\item [nCellsVirt] [{\tt integer}] number of cells in buildup,
	\item [nSamplVirt] [{\tt integer}] number of MC events per cell in build-up,
	\item [nBinVirt] [{\tt integer}] number of bins for search of the optimal division,
	\item [EvPerBinVirt] [{\tt integer}] maximum events (equiv.) per bin in buid-up,
	\item [MaxWtRejVirt] [{\tt integer}] maximum weight for rejection,
	\item [RanSeed] [{\tt integer}] pseudorandom generator seed.
\end{description}

Different contributions to the NLO cross sections (real, virtual) can differ from
each other by orders of magnitude. To improve relative and absolute
accuracy, a user can enlarge \textit{nCellsVirt}. In general, it also makes generation speed higher.

\subsubsection{Standard Model parameters}

\begin{description}
	\item [alpha, gf, alphas, conhc] [{\tt double}] a list of constants and coefficients: 
	\(\alpha_{EM}\), \(G_{\mu}\), \(\alpha_{S}\), conversion constant from GeV$^{-2}$ to pb,
	\item [mw, mz, mh] [{\tt double}] W, Z, Higgs boson masses,
	\item [wz, ww, wh, wtp] [{\tt double}] W, Z, Higgs and the top quark widths,
	\item [men, mel, mmn, mmo, mtn, mta] [{\tt double}] \(\nu_e,e,\nu_{\mu}, \mu, \nu_{\tau}, \tau\) lepton masses,
	\item [mdn, mup, mst, mch, mbt, mtp] [{\tt double}] \(d,u,s,c,b,t\) quark masses.
	
\end{description}

\subsection{Persistency}
{\tt ReneSANCe} can use {\tt FOAM} capability to save the constructed grid to 
disk and use it later for MC production skipping the exploration step. It is
convenient to use precomputed grid to run several instances of the {\tt ReneSANCe}
event generator in parallel to speed up the calculation. In this case, each
instance of {\tt ReneSANCe} must be initialized with a different pseudorandom number
generator seed.
To turn off exploration step, the \textit{explore[Born, Virt, Hard]} parameter must be set to false. In this case, {\tt ReneSANCe} tries to search for a grid file in the running
directory and starts generation in the case of success. In the case of
``all-in-one'' approach, only one grid is used  and
its building is controlled by the \textit{exploreHard} parameter.

\subsection{LR asymmetry}
The {\tt ReneSANCe} event generator has a special mode helping to get Left-Right asymmetry
\begin{equation*}
A_{LR}=\frac{\sigma_{RL}-\sigma_{LR}}{\sigma_{RL}+\sigma_{LR}},
\end{equation*}
obtained by comparing the differential cross section for longitudinally polarized initial beams. In the case when \textit{alr} parameter is set
to 1, events are generated according to the difference between cross sections $\sigma_{RL}-\sigma_{LR}$.
In the case when \textit{alr} parameter is set
to 2, events are generated according to the sum $\sigma_{RL}+\sigma_{LR}$.
After generation of both samples is completed, one can make a histogram for the LR asymmetry by
dividing the histograms for these special samples bin by bin.

\section{Numerical results}
\label{sec:validation}
In this section, we present numerical results obtained by means of {\tt ReneSANCe} for NLO EW corrections to the cross sections of the Bhabha and $e^+e^- \to ZH$ processes with different polarization of
initial particles.

For numerical validation we worked in the $\alpha(0)$-scheme and used the following set of input parameters \cite{Fleischer:2006ht}:
\begin{eqnarray}
&&\alpha^{-1}(0)= 137.03599976, \quad \Gamma_Z = 2.49977 \; \text{GeV},\nonumber\\
&&M_W = 80.4515 \; \text{GeV}, \quad M_Z = 91.1867 \; \text{GeV},\nonumber\\
&&M_H = 125 \; \text{GeV}, \quad m_e = 0.51099907 \; \text{MeV},\nonumber\\
&&m_\mu = 0.105658389 \; \text{GeV}, \quad m_\tau = 1.77705 \; \text{GeV},\nonumber\\
&&m_d = 0.083 \; \text{GeV}, \quad m_u = 0.062 \; \text{GeV},\nonumber\\
&&m_s = 0.215 \; \text{GeV}, \quad m_c = 1.5 \; \text{GeV},\nonumber\\
&&m_b = 4.7 \; \text{GeV}, \quad m_t = 173.8 \; \text{GeV}.
\end{eqnarray}

For the Bhabha process the scattering angles $\theta_{e^-}$ and $\theta_{e^+}$ of the final electron and positron were restricted by the conditions $|\cos{\theta_{e^-}}| < 0.9$ and
$|\cos{\theta_{e^+}}| < 0.9$. No restrictions were applied for 4-momenta of the final particles in the $e^+e^- \to ZH$ process.

The results for the Bhabha process are shown in Tables~\ref{eeee250}-\ref{eeee1000}, and the results for the $e^+e^- \to ZH$ process are shown in Tables~\ref{eezh250}-\ref{eezh1000} for center-of-mass
energy $\sqrt{s}$ values of 250 GeV, 500 GeV, and 1000 GeV. We present the numbers for the Born-level cross section $\sigma^{\text{Born}}$, the contribution of hard photon emission
$\sigma^{\text{hard}}$, the sum of Born, virtual and soft photon contributions $\sigma^{\text{B+v+s}}$, the total one-loop cross section $\sigma^{\text{1-loop}}$, and the relative correction $\delta$
that is defined as
\begin{equation}
\delta = \frac{\sigma^{\text{1-loop}} - \sigma^{\text{Born}}}{\sigma^{\text{Born}}}~100\%.
\end{equation}
The numbers are shown for 4 possible combinations of polarizations of the initial $e^+$ and $e^-$, assuming these particles to be $100\%$-polarized. The cross sections values for arbitrary longitudinal
polarization can be obtained with these numbers using the first row of Eq.~\ref{polxsec}.

We calculated $\sigma^{\text{1-loop}}$ using values $\omega = 10^{-6} \dfrac{\sqrt{s}}{2}$ and $\omega = 10^{-5} \dfrac{\sqrt{s}}{2}$ for the Bhabha process and $\omega = 10^{-5} \dfrac{\sqrt{s}}{2}$ and
$\omega = 10^{-4} \dfrac{\sqrt{s}}{2}$ for the $e^+e^- \to ZH$ process to check the stability of one-loop cross sections with respect to the choice of the photon energy that separates the phase space regions
associated with the soft and hard photon emission.

We compared our results for $\sigma^{\text{Born}}$ and $\sigma^{\text{hard}}$ with the corresponding numbers produced with the help of the {\tt CalcHEP} and {\tt WHIZARD} programs and obtained good
agreement. For this comparison we used a version of {\tt CalcHEP} with quadruple precision (16 bytes), provided to us by its authors. Calculation by {\tt WHIZARD} was performed with the extended precision
(10 bytes) option. Also, in the unpolarized case, we have agreement for $\sigma^{\text{B+v+s}}$ with the results of {\tt aITALC} \cite{Lorca:2004fg} for the Bhabha process and of {\tt Grace-Loop}
\cite{Belanger:2003sd} for the $e^+e^- \to ZH$ process.

As can be seen from the tables, the size of corrections strongly depends on the polarization of initial particles for both the processes.

One can notice that for the $e^+e^- \to ZH$ process the Born level contribution is zero in the case when the initial particles have $(-,-)$ or $(+,+)$ polarizations. In these cases, the only nonzero
contribution to the NLO EW cross section is due to the real emission that does not contain soft photon singularities for these polarizations and we can set $\omega = 0$.

\begin{table}[!h]
    \begin{center}
	\begin{tabular}{|r|r|l|l|l|l|l|l|l|}
	    \hline
	    \multirow{2}{*}{$P_{e^+}$} & \multirow{2}{*}{$P_{e^-}$} & \multirow{2}{*}{$\omega$, $\frac{\sqrt{s}}{2}$}
	    & \multirow{2}{*}{code} & \multirow{2}{*}{$\sigma^{\text{Born}}$, pb} & \multirow{2}{*}{$\sigma^{\text{hard}}$, pb}
	    & \multirow{2}{*}{$\sigma^{\text{B+v+s}}$, pb} & \multirow{2}{*}{$\sigma^{\text{1-loop}}$, pb} & \multirow{2}{*}{$\delta$, \%}\\
	    &&&&&&&&\\
	    \hline
	    \multirow{4}{*}{$-1$} & \multirow{4}{*}{$-1$} & $10^{-6}$ & \multirow{2}{*}{\tt ReneSANCe} & \multirow{2}{*}{55.263(1)} & 154.99(1) & $-93.396(1)$ & 61.60(1) & 11.46(1)\\
	    \cline{3-3}
	    \cline{6-9}
	    && \multirow{3}{*}{$10^{-5}$} &&& 127.65(1) & $-66.063(1)$ & 61.59(1) & 11.45(1)\\
	    \cline{4-9}
	    &&& {\tt WHIZARD} & 55.264(1) & 127.7(1) &&&\\
	    \cline{4-6}
	    &&& {\tt CalcHEP} & 55.263(1) & 127.6(1) &&&\\
	    \cline{3-9}
	    \hline
	    \multirow{4}{*}{$-1$} & \multirow{4}{*}{1} & $10^{-6}$ & \multirow{2}{*}{\tt ReneSANCe} & \multirow{2}{*}{55.346(1)} & 152.23(1) & $-92.551(1)$ & 59.69(1) & 7.83(2)\\
	    \cline{3-3}
	    \cline{6-9}
	    && \multirow{3}{*}{$10^{-5}$} &&& 125.11(1) & $-65.411(1)$ & 59.71(1) & 7.88(2)\\
	    \cline{4-9}
	    &&& {\tt WHIZARD} & 55.345(1) & 124.8(1) &&&\\
	    \cline{4-6}
	    &&& {\tt CalcHEP} & 55.346(1) & 125.2(1) &&&\\
	    \cline{3-9}
	    \hline
	    \multirow{4}{*}{1} & \multirow{4}{*}{$-1$} & $10^{-6}$ & \multirow{2}{*}{\tt ReneSANCe} & \multirow{2}{*}{60.834(1)} & 167.56(1) & $-103.633(2)$ & 63.92(1) & 5.08(1)\\
	    \cline{3-3}
	    \cline{6-9}
	    && \multirow{3}{*}{$10^{-5}$} &&& 137.71(1) & $-73.794(1)$ & 63.92(1) & 5.07(1)\\
	    \cline{4-9}
	    &&& {\tt WHIZARD} & 60.833(1) & 137.6(1) &&&\\
	    \cline{4-6}
	    &&& {\tt CalcHEP} & 60.834(1) & 137.8(1) &&&\\
	    \hline
	    \multirow{4}{*}{1} & \multirow{4}{*}{1} & $10^{-6}$ & \multirow{2}{*}{\tt ReneSANCe} & \multirow{2}{*}{55.253(1)} & 154.99(1) & $-93.396(1)$ & 61.60(1) & 11.46(1)\\
	    \cline{3-3}
	    \cline{6-9}
	    && \multirow{3}{*}{$10^{-5}$} &&& 127.66(1) & $-66.065(1)$ & 61.60(1) & 11.46(1)\\
	    \cline{4-9}
	    &&& {\tt WHIZARD} & 55.263(1) & 127.7(1) &&&\\
	    \cline{4-6}
	    &&& {\tt CalcHEP} & 55.263(1) & 127.6(1) &&&\\
	    \hline
	\end{tabular}
	\caption{{\tt ReneSANCe} results for the cross sections $\sigma^{\text{Born}}$, $\sigma^{\text{hard}}$, $\sigma^{\text{B+v+s}}$ and $\sigma^{\text{1-loop}}$ of the process $e^+e^- \to e^-e^+$ and
	relative correction $\delta$ for $\sqrt{s} = 250\text{ GeV}$ and various polarizations of the initial particles. For comparison, the numbers for $\sigma^{\text{Born}}$ and $\sigma^{\text{hard}}$
	obtained by means of {\tt CalcHEP} and {\tt WHIZARD} are given as well.}
	\label{eeee250}
    \end{center}
\end{table}

\begin{table}[!h]
    \begin{center}
	\begin{tabular}{|r|r|l|l|l|l|l|l|l|}
	    \hline
	    \multirow{2}{*}{$P_{e^+}$} & \multirow{2}{*}{$P_{e^-}$} & \multirow{2}{*}{$\omega$, $\frac{\sqrt{s}}{2}$}
	    & \multirow{2}{*}{code} & \multirow{2}{*}{$\sigma^{\text{Born}}$, pb} & \multirow{2}{*}{$\sigma^{\text{hard}}$, pb}
	    & \multirow{2}{*}{$\sigma^{\text{B+v+s}}$, pb} & \multirow{2}{*}{$\sigma^{\text{1-loop}}$, pb} & \multirow{2}{*}{$\delta$, \%}\\
	    &&&&&&&&\\
	    \hline
	    \multirow{4}{*}{$-1$} & \multirow{4}{*}{$-1$} & $10^{-6}$ & \multirow{2}{*}{\tt ReneSANCe} & \multirow{2}{*}{10.554(1)} & 31.406(1) & $-19.496(1)$ & 11.910(1) & 12.84(1)\\
	    \cline{3-3}
	    \cline{6-9}
	    && \multirow{3}{*}{$10^{-5}$} &&& 25.866(1) & $-13.956(1)$ & 11.910(1) & 12.85(1)\\
	    \cline{4-9}
	    &&& {\tt WHIZARD} & 10.554(1) & 25.87(1) &&&\\
	    \cline{4-6}
	    &&& {\tt CalcHEP} & 10.554(1) & 25.86(1) &&&\\
	    \cline{3-9}
	    \hline
	    \multirow{4}{*}{$-1$} & \multirow{4}{*}{1} & $10^{-6}$ & \multirow{2}{*}{\tt ReneSANCe} & \multirow{2}{*}{16.574(1)} & 48.166(3) & $-30.183(1)$ & 17.983(3) & 8.50(2)\\
	    \cline{3-3}
	    \cline{6-9}
	    && \multirow{3}{*}{$10^{-5}$} &&& 39.547(2) & $-21.565(1)$ & 17.982(2) & 8.49(1)\\
	    \cline{4-9}
	    &&& {\tt WHIZARD} & 16.575(1) & 39.52(1) &&&\\
	    \cline{4-6}
	    &&& {\tt CalcHEP} & 16.574(1) & 39.49(1) &&&\\
	    \cline{3-9}
	    \hline
	    \multirow{4}{*}{1} & \multirow{4}{*}{$-1$} & $10^{-6}$ & \multirow{2}{*}{\tt ReneSANCe} & \multirow{2}{*}{19.832(1)} & 57.618(3) & $-37.594(1)$ & 20.024(3) & 0.96(2)\\
	    \cline{3-3}
	    \cline{6-9}
	    && \multirow{3}{*}{$10^{-5}$} &&& 47.300(3) & $-27.278(2)$ & 20.022(3) & 0.95(1)\\
	    \cline{4-9}
	    &&& {\tt WHIZARD} & 19.832(1) & 47.27(1) &&&\\
	    \cline{4-6}
	    &&& {\tt CalcHEP} & 19.833(1) & 47.26(1) &&&\\
	    \hline
	    \multirow{4}{*}{1} & \multirow{4}{*}{1} & $10^{-6}$ & \multirow{2}{*}{\tt ReneSANCe} & \multirow{2}{*}{10.554(1)} & 31.407(1) & $-19.497(1)$ & 11.910(1) & 12.85(1)\\
	    \cline{3-3}
	    \cline{6-9}
	    && \multirow{3}{*}{$10^{-5}$} &&& 25.865(1) & $-13.956(1)$ & 11.909(1) & 12.84(1)\\
	    \cline{4-9}
	    &&& {\tt WHIZARD} & 10.554(1) & 25.86(1) &&&\\
	    \cline{4-6}
	    &&& {\tt CalcHEP} & 10.554(1) & 25.86(1) &&&\\
	    \hline
	\end{tabular}
	\caption{{\tt ReneSANCe} results for the cross sections $\sigma^{\text{Born}}$, $\sigma^{\text{hard}}$, $\sigma^{\text{B+v+s}}$ and $\sigma^{\text{1-loop}}$ of the process $e^+e^- \to e^-e^+$ and
	relative correction $\delta$ for $\sqrt{s} = 500\text{ GeV}$ and various polarizations of the initial particles. For comparison, the numbers for $\sigma^{\text{Born}}$ and $\sigma^{\text{hard}}$
	obtained by means of {\tt CalcHEP} and {\tt WHIZARD} are given as well.}
	\label{eeee500}
    \end{center}
\end{table}

\begin{table}[!h]
    \begin{center}
	\begin{tabular}{|r|r|l|l|l|l|l|l|l|}
	    \hline
	    \multirow{2}{*}{$P_{e^+}$} & \multirow{2}{*}{$P_{e^-}$} & \multirow{2}{*}{$\omega$, $\frac{\sqrt{s}}{2}$}
	    & \multirow{2}{*}{code} & \multirow{2}{*}{$\sigma^{\text{Born}}$, pb} & \multirow{2}{*}{$\sigma^{\text{hard}}$, pb}
	    & \multirow{2}{*}{$\sigma^{\text{B+v+s}}$, pb} & \multirow{2}{*}{$\sigma^{\text{1-loop}}$, pb} & \multirow{2}{*}{$\delta$, \%}\\
	    &&&&&&&&\\
	    \hline
	    \multirow{4}{*}{$-1$} & \multirow{4}{*}{$-1$} & $10^{-6}$ & \multirow{2}{*}{\tt ReneSANCe} & \multirow{2}{*}{2.2208(1)} & 6.9922(2) & $-4.4841(1)$ & 2.5081(2) & 12.94(1)\\
	    \cline{3-3}
	    \cline{6-9}
	    && \multirow{3}{*}{$10^{-5}$} &&& 5.7579(1) & $-3.2500(1)$ & 2.5080(1) & 12.93(1)\\
	    \cline{4-9}
	    &&& {\tt WHIZARD} & 2.2208(1) & 5.759(1) &&&\\
	    \cline{4-6}
	    &&& {\tt CalcHEP} & 2.2208(1) & 5.758(1) &&&\\
	    \cline{3-9}
	    \hline
	    \multirow{4}{*}{$-1$} & \multirow{4}{*}{1} & $10^{-6}$ & \multirow{2}{*}{\tt ReneSANCe} & \multirow{2}{*}{4.5715(1)} & 14.040(1) & $-9.0239(1)$ & 5.017(1) & 9.74(2)\\
	    \cline{3-3}
	    \cline{6-9}
	    && \multirow{3}{*}{$10^{-5}$} &&& 11.532(1) & $-6.5125(1)$ & 5.019(1) & 9.79(3)\\
	    \cline{4-9}
	    &&& {\tt WHIZARD} & 4.5715(1) & 11.53(1) &&&\\
	    \cline{4-6}
	    &&& {\tt CalcHEP} & 4.5715(1) & 11.52(1) &&&\\
	    \cline{3-9}
	    \hline
	    \multirow{4}{*}{1} & \multirow{4}{*}{$-1$} & $10^{-6}$ & \multirow{2}{*}{\tt ReneSANCe} & \multirow{2}{*}{5.7038(1)} & 17.512(1) & $-12.099(1)$ & 5.412(1) & $-5.11(2)$\\
	    \cline{3-3}
	    \cline{6-9}
	    && \multirow{3}{*}{$10^{-5}$} &&& 14.381(1) & $-8.9659(1)$ & 5.416(1) & $-5.05(2)$\\
	    \cline{4-9}
	    &&& {\tt WHIZARD} & 5.7038(1) & 14.37(1) &&&\\
	    \cline{4-6}
	    &&& {\tt CalcHEP} & 5.7038(1) & 14.36(1) &&&\\
	    \hline
	    \multirow{4}{*}{1} & \multirow{4}{*}{1} & $10^{-6}$ & \multirow{2}{*}{\tt ReneSANCe} & \multirow{2}{*}{2.2208(1)} & 6.9920(2) & $-4.4842(1)$ & 2.5078(2) & 12.92(1)\\
	    \cline{3-3}
	    \cline{6-9}
	    && \multirow{3}{*}{$10^{-5}$} &&& 5.7579(1) & $-3.2500(1)$ & 2.5079(1) & 12.93(1)\\
	    \cline{4-9}
	    &&& {\tt WHIZARD} & 2.2208(1) & 5.758(1) &&&\\
	    \cline{4-6}
	    &&& {\tt CalcHEP} & 2.2208(1) & 5.759(1) &&&\\
	    \hline
	\end{tabular}
	\caption{{\tt ReneSANCe} results for the cross sections $\sigma^{\text{Born}}$, $\sigma^{\text{hard}}$, $\sigma^{\text{B+v+s}}$ and $\sigma^{\text{1-loop}}$ of the process $e^+e^- \to e^-e^+$ and
	relative correction $\delta$ for $\sqrt{s} = 1000\text{ GeV}$ and various polarizations of the initial particles. For comparison, the numbers for $\sigma^{\text{Born}}$ and $\sigma^{\text{hard}}$
	obtained by means of {\tt CalcHEP} and {\tt WHIZARD} are given as well.}
	\label{eeee1000}
    \end{center}
\end{table}

\begin{table}[!ht]
    \begin{center}
	\begin{tabular}{|r|r|l|l|l|l|l|l|l|}
	    \hline
	    \multirow{2}{*}{$P_{e^+}$} & \multirow{2}{*}{$P_{e^-}$} & \multirow{2}{*}{$\omega$, $\frac{\sqrt{s}}{2}$}
	    & \multirow{2}{*}{code} & \multirow{2}{*}{$\sigma^{\text{Born}}$, fb} & \multirow{2}{*}{$\sigma^{\text{hard}}$, fb}
	    & \multirow{2}{*}{$\sigma^{\text{B+v+s}}$, fb} & \multirow{2}{*}{$\sigma^{\text{1-loop}}$, fb} & \multirow{2}{*}{$\delta$, \%}\\
	    &&&&&&&&\\
	    \hline
	    \multirow{3}{*}{$-1$} & \multirow{3}{*}{$-1$} & \multirow{3}{*}{0} & {\tt ReneSANCe} & 0 & 0.0260(1) & 0 & 0.0260(1) & $+\infty$\\
	    \cline{4-9}
	    &&& {\tt WHIZARD} & 0 & 0.0259(1) &&&\\
	    \cline{4-6}
	    &&& {\tt CalcHEP} & 0 & 0.0260(1) &&&\\
	    \hline
	    \multirow{4}{*}{$-1$} & \multirow{4}{*}{1} & $10^{-5}$ & \multirow{2}{*}{\tt ReneSANCe} & \multirow{2}{*}{350.00(1)} & 400.85(1) & $-28.82(1)$ & 372.03(1) & 6.30(1)\\
	    \cline{3-3}
	    \cline{6-9}
	    && \multirow{3}{*}{$10^{-4}$} &&& 306.51(1) & 65.53(1) & 372.04(1) & 6.30(1)\\
	    \cline{4-9}
	    &&& {\tt WHIZARD} & 349.99(1) & 306.6(2) &&&\\
	    \cline{4-6}
	    &&& {\tt CalcHEP} & 350.00(1) & 306.5(1) &&&\\
	    \cline{3-9}
	    \hline
	    \multirow{4}{*}{1} & \multirow{4}{*}{$-1$} & $10^{-5}$ & \multirow{2}{*}{\tt ReneSANCe} & \multirow{2}{*}{552.45(1)} & 632.74(1) & $-177.74(1)$ & 455.00(1) & $-17.64(1)$\\
	    \cline{3-3}
	    \cline{6-9}
	    && \multirow{3}{*}{$10^{-4}$} &&& 483.80(1) & $-28.81(1)$ & 454.99(1) & $-17.64(1)$\\
	    \cline{4-9}
	    &&& {\tt WHIZARD} & 552.45(1) & 483.7(3) &&&\\
	    \cline{4-6}
	    &&& {\tt CalcHEP} & 552.46(1) & 483.7(1) &&&\\
	    \hline
	    \multirow{3}{*}{1} & \multirow{3}{*}{1} & \multirow{3}{*}{0} & {\tt ReneSANCe} & 0 & 0.0260(1) & 0 & 0.0260(1) & $+\infty$\\
	    \cline{4-9}
	    &&& {\tt WHIZARD} & 0 & 0.0260(1) &&&\\
	    \cline{4-6}
	    &&& {\tt CalcHEP} & 0 & 0.0261(1) &&&\\
	    \hline
	\end{tabular}
	\caption{{\tt ReneSANCe} results for the cross sections $\sigma^{\text{Born}}$, $\sigma^{\text{hard}}$, $\sigma^{\text{B+v+s}}$ and $\sigma^{\text{1-loop}}$ of the process $e^+e^- \to ZH$ and
	relative correction $\delta$ for $\sqrt{s} = 250\text{ GeV}$ and various polarizations of the initial particles. For comparison, the numbers for $\sigma^{\text{Born}}$ and $\sigma^{\text{hard}}$
	obtained by means of {\tt CalcHEP} and {\tt WHIZARD} are given as well.}
	\label{eezh250}
    \end{center}
\end{table}

\begin{table}[!h]
    \begin{center}
	\begin{tabular}{|r|r|l|l|l|l|l|l|l|}
	    \hline
	    \multirow{2}{*}{$P_{e^+}$} & \multirow{2}{*}{$P_{e^-}$} & \multirow{2}{*}{$\omega$, $\frac{\sqrt{s}}{2}$}
	    & \multirow{2}{*}{code} & \multirow{2}{*}{$\sigma^{\text{Born}}$, fb} & \multirow{2}{*}{$\sigma^{\text{hard}}$, fb}
	    & \multirow{2}{*}{$\sigma^{\text{B+v+s}}$, fb} & \multirow{2}{*}{$\sigma^{\text{1-loop}}$, fb} & \multirow{2}{*}{$\delta$, \%}\\
	    &&&&&&&&\\
	    \hline
	    \multirow{3}{*}{$-1$} & \multirow{3}{*}{$-1$} & \multirow{3}{*}{0} & {\tt ReneSANCe} & 0 & 0.2199(1) & 0 & 0.2199(1) & $+\infty$\\
	    \cline{4-9}
	    &&& {\tt WHIZARD} & 0 & 0.2200(1) &&&\\
	    \cline{4-6}
	    &&& {\tt CalcHEP} & 0 & 0.2200(1) &&&\\
	    \hline
	    \multirow{4}{*}{$-1$} & \multirow{4}{*}{1} & $10^{-5}$ & \multirow{2}{*}{\tt ReneSANCe} & \multirow{2}{*}{83.373(1)} & 121.97(1) & $-12.05(1)$ & 109.92(1) & 31.84(1)\\
	    \cline{3-3}
	    \cline{6-9}
	    && \multirow{3}{*}{$10^{-4}$} &&& 98.26(1) & 11.66(1) & 109.92(1) & 31.84(1)\\
	    \cline{4-9}
	    &&& {\tt WHIZARD} & 83.373(1) & 98.34(1) &&&\\
	    \cline{4-6}
	    &&& {\tt CalcHEP} & 83.373(1) & 98.27(1) &&&\\
	    \cline{3-9}
	    \hline
	    \multirow{4}{*}{1} & \multirow{4}{*}{$-1$} & $10^{-5}$ & \multirow{2}{*}{\tt ReneSANCe} & \multirow{2}{*}{131.60(1)} & 192.53(1) & $-53.16(1)$ & 139.37(1) & 5.91(1)\\
	    \cline{3-3}
	    \cline{6-9}
	    && \multirow{3}{*}{$10^{-4}$} &&& 155.10(1) & $-15.73(1)$ & 139.37(1) & 5.90(1)\\
	    \cline{4-9}
	    &&& {\tt WHIZARD} & 131.60(1) & 155.23(1) &&&\\
	    \cline{4-6}
	    &&& {\tt CalcHEP} & 131.60(1) & 155.08(1) &&&\\
	    \hline
	    \multirow{3}{*}{1} & \multirow{3}{*}{1} & \multirow{3}{*}{0} & {\tt ReneSANCe} & 0 & 0.2200(1) & 0 & 0.2200(1) & $+\infty$\\
	    \cline{4-9}
	    &&& {\tt WHIZARD} & 0 & 0.2200(1) &&&\\
	    \cline{4-6}
	    &&& {\tt CalcHEP} & 0 & 0.2201(1) &&&\\
	    \hline
	\end{tabular}
	\caption{{\tt ReneSANCe} results for the cross sections $\sigma^{\text{Born}}$, $\sigma^{\text{hard}}$, $\sigma^{\text{B+v+s}}$ and $\sigma^{\text{1-loop}}$ of the process $e^+e^- \to ZH$ and
	relative correction $\delta$ for $\sqrt{s} = 500\text{ GeV}$ and various polarizations of the initial particles. For comparison, the numbers for $\sigma^{\text{Born}}$ and $\sigma^{\text{hard}}$
	obtained by means of {\tt CalcHEP} and {\tt WHIZARD} are given as well.}
	\label{eezh500}
    \end{center}
\end{table}

\begin{table}[!h]
    \begin{center}
	\begin{tabular}{|r|r|l|l|l|l|l|l|l|}
	    \hline
	    \multirow{2}{*}{$P_{e^+}$} & \multirow{2}{*}{$P_{e^-}$} & \multirow{2}{*}{$\omega$, $\frac{\sqrt{s}}{2}$}
	    & \multirow{2}{*}{code} & \multirow{2}{*}{$\sigma^{\text{Born}}$, fb} & \multirow{2}{*}{$\sigma^{\text{hard}}$, fb}
	    & \multirow{2}{*}{$\sigma^{\text{B+v+s}}$, fb} & \multirow{2}{*}{$\sigma^{\text{1-loop}}$, fb} & \multirow{2}{*}{$\delta$, \%}\\
	    &&&&&&&&\\
	    \hline
	    \multirow{3}{*}{$-1$} & \multirow{3}{*}{$-1$} & \multirow{3}{*}{0} & {\tt ReneSANCe} & 0 & 0.1327(1) & 0 & 0.1327(1) & $+\infty$\\
	    \cline{4-9}
	    &&& {\tt WHIZARD} & 0 & 0.1327(1) &&&\\
	    \cline{4-6}
	    &&& {\tt CalcHEP} & 0 & 0.1329(1) &&&\\
	    \hline
	    \multirow{4}{*}{$-1$} & \multirow{4}{*}{1} & $10^{-5}$ & \multirow{2}{*}{\tt ReneSANCe} & \multirow{2}{*}{18.702(1)} & 30.874(1) & $-4.979(1)$ & 25.895(1) & 38.46(1)\\
	    \cline{3-3}
	    \cline{6-9}
	    && \multirow{3}{*}{$10^{-4}$} &&& 25.280(1) & 0.617(1) & 25.897(1) & 38.47(1)\\
	    \cline{4-9}
	    &&& {\tt WHIZARD} & 18.702(1) & 25.26(1) &&&\\
	    \cline{4-6}
	    &&& {\tt CalcHEP} & 18.702(1) & 25.281(2) &&&\\
	    \cline{3-9}
	    \hline
	    \multirow{4}{*}{1} & \multirow{4}{*}{$-1$} & $10^{-5}$ & \multirow{2}{*}{\tt ReneSANCe} & \multirow{2}{*}{29.521(1)} & 48.734(2) & $-16.644(1)$ & 32.090(2) & 8.70(1)\\
	    \cline{3-3}
	    \cline{6-9}
	    && \multirow{3}{*}{$10^{-4}$} &&& 39.904(1) & $-7.810(1)$ & 32.094(1) & 8.72(1)\\
	    \cline{4-9}
	    &&& {\tt WHIZARD} & 29.521(1) & 39.93(2) &&&\\
	    \cline{4-6}
	    &&& {\tt CalcHEP} & 29.521(1) & 39.900(4) &&&\\
	    \hline
	    \multirow{3}{*}{1} & \multirow{3}{*}{1} & \multirow{3}{*}{0} & {\tt ReneSANCe} & 0 & 0.1327(1) & 0 & 0.1327(1) & $+\infty$\\
	    \cline{4-9}
	    &&& {\tt WHIZARD} & 0 & 0.1326(1) &&&\\
	    \cline{4-6}
	    &&& {\tt CalcHEP} & 0 & 0.1327(1) &&&\\
	    \hline
	\end{tabular}
	\caption{{\tt ReneSANCe} results for the cross sections $\sigma^{\text{Born}}$, $\sigma^{\text{hard}}$, $\sigma^{\text{B+v+s}}$ and $\sigma^{\text{1-loop}}$ of the process $e^+e^- \to ZH$ and
	relative correction $\delta$ for $\sqrt{s} = 1000\text{ GeV}$ and various polarizations of the initial particles. For comparison, the numbers for $\sigma^{\text{Born}}$ and $\sigma^{\text{hard}}$
	obtained by means of {\tt CalcHEP} and {\tt WHIZARD} are given as well.}
	\label{eezh1000}
    \end{center}
\end{table}

\section{Summary}
\label{sec:summary}
The {\tt ReneSANCe} Monte Carlo event generator is a new tool for studying processes at high-energy $e^+e^-$ colliders. It can produce unweighted events at the one-loop EW precision level taking into account
polarization of the initial particles. The generated events can be saved in {\tt LHEF} and {\tt ROOT} format.

\section{Acknowledgements}
\label{sec:acknowledgements}
The authors are grateful to A.~Arbuzov, S.~Bondarenko, I.~Boyko, Ya.~Dydyshka, L.~Kalinovskaya, L.~Rumyantsev and Yu.~Yermolchyk for their help in developing and debugging the code and valuable comments in writing the paper.
We thank A.~Belyaev and A.~Pukhov for providing a version of {\tt CalcHEP} program that includes the polarization of massive particles with quadruple precision, which allowed us to perform the
relevant cross-checks.

This work was supported by the RFBR grant 20-02-00441.

\bibliographystyle{elsarticle-num}
\bibliography{CPC_ReneSANCe}
\end{document}